# 3D-imaging of Printed Nanostructured Networks using High-resolution FIB-SEM Nanotomography

Cian Gabbett, Luke Doolan, Kevin Synnatschke, Laura Gambini, Emmet Coleman, Adam G. Kelly, Shixin Liu, Eoin Caffrey, Jose Munuera, Catriona Murphy, Stefano Sanvito, Lewys Jones, Jonathan N. Coleman*

School of Physics, CRANN and AMBER Research Centres, Trinity College Dublin, Dublin 2, Ireland*

*colemaj@tcd.ie

## Abstract

Networks of solution-processed nanomaterials are important for multiple applications in electronics, sensing and energy storage/generation. While it is known that network morphology plays a dominant role in determining the physical properties of printed networks, it remains difficult to quantify network structure. Here, we utilise FIB-SEM nanotomography to characterise the morphology of nanostructured networks. Nanometer-resolution 3D-images were obtained from printed networks of graphene nanosheets of various sizes, as well as networks of $WS_2$ nanosheets, silver nanosheets and silver nanowires. Important morphological characteristics, including network porosity, tortuosity, pore dimensions and nanosheet orientation were extracted and linked to network resistivity. By extending this technique to interrogate the structure and interfaces within vertical printed heterostacks, we demonstrate the potential of this technique for device characterisation and optimisation.

## Introduction

Liquid-deposited networks of 0D nanoparticles, 1D nanowires or nanotubes and 2D nanosheets have shown great promise for applications in areas from printed electronics to sensing to energy storage (*1-4*). In particular, devices based on networks of 2D materials have been the subject of intensive research (*5, 6*), largely due to the very broad range of such materials (*7*), as well as recent advances in both scalable nanosheet production and deposition techniques (*8*). An array of devices based on printed nanosheet networks have been demonstrated including supercapacitors (*9*), transistors (*10*) and strain-sensors (*11*). Recently, it has become clear that the performance of such devices is very sensitive to the morphology of the nanosheet network (*12*). Printed networks tend to consist of somewhat porous, disordered arrays of nanosheets with variable degrees of connectivity, alignment and inter-sheet coupling. However, the morphology of such networks has proven difficult to characterise quantitatively.

While standard techniques such as mercury intrusion porosimetry (MIP) and $N_2$ BET analysis have been used to determine the pore-size-distribution and specific surface area in thick nanosheet networks (*13*), such methods require prohibitively large sample volumes (i.e. thick films), and considerable sample preparation (*14*). Alternatively, AFM and SEM only provide surface information, although SEM can also analyse cross-sections (*15*). While 3D-imaging would be ideal, X-ray computed tomography (X-ray CT) is resolution-limited by voxel sizes of 30 - 500 nm (*16, 17*), which are not sufficient to quantify the structure of nanomaterial networks (*18, 19*).

Here, we present FIB-SEM nanotomography (FIB-SEM-NT) as a viable means to interrogate the morphology of nanostructured networks (*20*). We report 3D-imaging with a voxel size of 5 nm × 5 nm × 15 nm and demonstrate the extraction of quantitative morphological information from these images. Finally, we demonstrate a machine-learning protocol to generate cubic voxels.

**Results and Discussion**

*3D-imaging using FIB-SEM Nanotomography*

Liquid phase exfoliation (LPE) yields suspensions of nanosheets in various solvents (Fig. 1A) which can then be printed into networks (*21, 22*). Conventional surface and cross-sectional SEM imaging of a spray-cast graphene nanosheet network (nanosheet length, $l_{NS}$ = 695 nm, Fig. 1B) allows qualitative observations of network porosity or nanosheet alignment. However, it is difficult to extract quantitative information. Here, we use FIB-SEM-NT to produce a high-resolution 3D reconstruction of portions of the network, which we refer to as 3D-Imaging. To achieve this (Fig. 1C), ~800 network cross-sections were sequentially milled using the FIB and imaged using the SEM (*23*). Each network slice has an in-plane pixel size of 5 nm and average thickness of ~ 15 nm, giving a voxel size of ~ 375 $nm^3$, 10 – 1000 times smaller than conventional X-ray CT scanners (*24, 25*). A typical network volume of 20 μm × 15 μm × 2 μm contains $1.6×10^9$ voxels.

To enable quantitative analysis, each image in the stack was classified into its pore and nanosheet components using trainable *WEKA* segmentation (*26*). As shown in Fig. 1D, regions of each slice were first manually assigned as either nanosheet or pore, providing training data for the classifier. Each slice was then segmented into these components on a pixel-by-pixel basis and converted to a binary image containing only nanosheets or pores. These segmented image stacks were then aligned and interpolated in *Dragonfly* to form a 3D network reconstruction (Fig. 1E). This allows the morphological properties of the network volume to be analysed on a voxel-by-voxel basis at a resolution that preserves the discrete nanosheet/pore components, as shown in Fig. 1E.

*Analysing 3D-images of nanosheet networks*

A graphene network ($l_{NS}$ = 238 nm), split into its pore and nanosheet volumes, is shown in Fig. 2A. The network porosity, P, was measured both across the entire volume and on a slice-by-slice basis (Fig. 2B) (*27, 28*). While the global porosity was 41 ± 1%, the scatter in the slice-by-slice data highlights the local inhomogeneity of the network.

The pore connectivity is a key parameter that determines the accessible surface area in sensing or electrochemical applications (*29, 30*). The pore volume (Fig. 2A) was found to be highly contiguous, consistent with MIP/BET data for filtered graphene networks (*13*), with >95% of the total pore volume contained in a single macropore with a connectivity length-scale of ~250 nm (Fig. 2C) (*28*).

To quantitatively assess the pore and nanosheet connectivity we calculated the tortuosity factor (*31*), κ, by determining the reduction of diffusive flux through each network in the in-plane (IP, *x, z*) and out-of-plane (OOP, *y*) directions using *TauFactor* (*27*) (Fig. 2D). Bottlenecks in the diffusive flux through the pore volume are shown visually in Fig. 2D. The comparatively smaller nanosheet κ-values are reflective of a well-connected nanosheet network while $κ_{OOP}/κ_{IP}>1$ suggests that the nanosheets are primarily aligned in-plane. This is consistent with directional anisotropies in conductivity and mass transport through solution-processed 2D networks (*32*).

Because the pore volume is highly connected, we analyse the cross-sectional area and shape of 2D pore sections in each slice. The pore circularity, C, is plotted as a function of pore cross-sectional area, A, for each pore in the network in Fig. 2E ($C = 4\pi A/(perimeter)^2$, smaller values mean more elongated pores). The area-weighted heat map suggests that the pore volume is dominated by pore chambers with $A>10^4$ nm$^2$, and that larger pores have elongated geometries, consistent with published BET measurements (*13*).

By isolating discrete nanoplatelets and noting their orientation (Fig. 2F Inset) (*33*), the distribution of angles, φ, between each nanoplatelet normal vector and the y-direction was calculated (Fig. 2F). The Hermans orientation factor (*34*), $S = (3\langle\cos^2\varphi\rangle - 1)/2$, was 0.61 ± 0.07 for the network, consistent with partial in-plane alignment (S=1 for perfect alignment). This is in broad agreement with a value of S = 0.79 for a vacuum filtered $Ti_3C_2T_x$ nanosheet network measured using wide angle X-ray scattering (WAXS) (*16*).

*Characterising size-dependent morphology*

The physical properties of 2D networks are known to scale with nanosheet size (*35, 36*). Here, we use FIB-SEM-NT to systematically study the morphology of LPE graphene networks for various nanosheet lengths, $l_{NS}$. Size-selected inks were produced using liquid cascade centrifugation (*37*), characterised by AFM (Fig. 3A) and sprayed into networks. Representative FIB-SEM cross-sections (Fig. 3B) show noticeable changes in network morphology as $l_{NS}$ is increased from 80 to 1087 nm. Analysis shows a clear increase in network porosity from 39 to 51 % with increasing $l_{NS}$ (Fig. 3C), with a corresponding increase in the characteristic pore size, $\zeta = \sqrt{\langle A \rangle}$ (Fig. 3D).

The dependence of network morphology on $l_{NS}$ is further reflected in the specific surface area (SSA) of the network (Fig. 3E), which decreased from 23 to 15 m$^2$.g$^{-1}$ as $l_{NS}$ increased from 80 to 1087 nm. This is consistent with expectations that SSA scales inversely with nanosheet thickness (*38*), $t_{NS}$, given $t_{NS}$ and $l_{NS}$ are intrinsically coupled for LPE (*39*), and agrees with a value of 25 m$^2$.g$^{-1}$ for $V_2O_5$ nanosheet films measured using BET (*40*). Such low values of SSA (*cf.* 2600 m$^2$.g$^{-1}$ for monolayers (*41*)) are due to restacking during deposition (*42*). By comparing the dimensions of the aggregated nanosheets in each network ($t_{Net}$, $l_{Net}$) to the dimensions measured by AFM in the inks, we found $l_{Net}/l_{NS}\sim 1$, while $t_{Net}/t_{NS} \gg 1$. This suggests that nanosheets restack with maximised basal plane overlap in these networks, with an apparent increase in aggregation observed for smaller $l_{NS}$.

The tortuosity factor, κ, of both the pore and nanosheet volumes was calculated as a function of $l_{NS}$. Both nanosheet and pore network tortuosities are significantly larger in the out-of-plane direction due to nanosheet in-plane alignment. Interestingly, the pore volume was found to become less tortuous with increasing $l_{NS}$, with the nanosheet network showing the opposite trend. This is because the tortuosity of a given network is known to scale with the fractional volume occupied by that network (*43*). This leads to a well-defined relationship between tortuosity and network volume fraction ($V_f$) for both pore ($V_{f,P}=P$) and nanosheet ($V_{f,NS}=1-P$) networks (Fig. 3F). The data follow an adjusted Bruggeman relation (*44*), $\kappa = \alpha V_f^{1-\beta}$. The extracted exponents for pores ($\beta_{IP}$ = 3.3 and $\beta_{OOP}$ = 5.6) and nanosheets ($\beta_{IP}$ = 1.9 and $\beta_{OOP}$ = 2.5) are considerably larger than values of 1.5 predicted by basic models (*43*). However, experimentally measured exponents (*45*) are usually >1.5, with values of $\beta$ = 2-5 predicted for high-aspect ratio particles (*46*).

To characterise the nanosheet orientation, we calculated the Hermans orientation factor, S, finding values of 0.54 to 0.7 (Fig. 3G), again implying in-plane nanosheet alignment. This is in agreement with values of S = 0.67 – 0.87 reported for vacuum filtered GO networks measured using small-angle X-ray scattering (SAXS) (*47*). Interestingly, the orientation factor appears to increase with increasing nanosheet length, $l_{NS}$ (Fig. 3G). This contrasts with solution-processed GO networks, where increased aspect ratios are known to drive improved alignment (*35, 48*). However, LPE nanosheets are comparatively smaller with roughly constant aspect ratios driven by nanosheet mechanics (*39*). Thus, we propose that the smaller, thinner nanosheets conform to each other more easily (*49*), leading to a reduction in network porosity while increasing the angular dispersion.

The network morphology should strongly impact physical properties. It can be shown that the in-plane resistivity of a nanosheet network, $\rho_{IP}$, is dependent on P, $l_{NS}$ and $\kappa_{IP}$, as well as the junction resistance, $R_J$, nanosheet resistivity, $\rho_{NS}$, and aspect ratio, $k_{NS}$:

$$\rho_{IP} \approx \frac{\rho_{NS} + (R_J l_{NS} / k_{NS})}{(1-P)/\kappa_{IP}} \qquad \text{Eqn.1}$$

We measured the network resistivity as a function of $l_{NS}$, with the data plotted in a linearized form in Fig. 3H. We find a very good straight line fit and reasonable values (*12*) of $R_J$ ~ 3 k$\Omega$ and $\rho_{NS}$ ~ 31 µ$\Omega$.m, highlighting the dependence of network resistivity on morphological factors.

*Versatility of 3D-imaging*

FIB-SEM nanotomography is a general method to analyse nanoscale networks. To show this, we 3D-imaged printed networks of $WS_2$ nanosheets, silver nanosheets (AgNS) and silver nanowires (AgNWs) (Fig. 4A-C). These images yield unprecedented insights into the structure of these networks. While there are distinct morphological differences between these systems, there are also similarities. Increasing nanosheet/nanowire length caused the network porosity and pore tortuosity to increase and decrease respectively (Fig. 4D-F), similar to graphene networks.

However, the magnitude of the porosity change was somewhat smaller for AgNS than in $WS_2$ (or graphene) networks. These differences in porosity scaling with $l_{NS}$ suggest that there are material-dependent factors contributing to the network morphology. Indeed, while the ratio of ($\kappa_{OOP}/\kappa_{IP}$) for the AgNS and $WS_2$ nanosheet networks was found to be ~ 1.5 and ~ 2.5 respectively, inferring considerable in-plane alignment, they are each different from the value of ~ 2.9 found for graphene networks.

The aligned restacking, previously discussed for graphene, can clearly be seen in AgNS networks, where vertical stacks of 3 - 7 AgNS are visible (Fig. 4B, Inset). A similar effect can be inferred for the $WS_2$ nanosheets.

The 3D-image of the AgNW network (Fig. 4C) allows AgNWs to be resolved both as isolated 1D objects and within small bundles. This highlights the resolution advantage of FIB-SEM-NT, as the AgNW diameter of ~ 55 nm is smaller than the pixel size for many X-ray CT techniques (50). As with the nanosheet data, the porosity of an AgNW network is seen to increase with increasing AgNW length in a manner consistent with a modified version of a reported model (51) (Fig. 4F). The ($\kappa_{OOP}/\kappa_{IP}$) ratio of ~ 1 in the AgNW networks is driven by the considerably higher porosities of P = 82 – 91%.

*Characterising electrical and device properties*

It has been reported that networks of electrochemically exfoliated (EE) nanosheets (52) display much higher mobility than their LPE counterparts (10) for reasons of morphology (12). Here, we utilise FIB-SEM-NT to investigate the morphological differences between EE ($t_{NS}$=4 nm) and LPE graphene nanosheet ($t_{NS}$=20 nm) networks deposited under identical conditions. 3D-imaging shows some differences in P and $\kappa_{IP}$ while electrical measurements show the EE network to be ~6 times more conductive (Fig. 5A). Combining the values in Fig. 5A with Eqn. 1 (using $t_{NS} = l_{NS}/k_{NS}$) and approximating these networks as purely junction limited ($R_J \gg R_{NS} = \rho_{NS}/t_{NS}$), implies $R_J$ in EE and LPE networks to be very similar. This means that against expectations, the conductivity difference in this case is predominately due to the nanosheet thickness difference rather than the morphological factors (P, $\kappa_{IP}$ and $R_J$).

By reconstructing the surface topography of both networks (Fig. 5A) we measured the root mean square roughness, $R_{RMS}$, of the EE and LPE networks to be ~122 nm and ~182 nm respectively. The reduced surface roughness in the EE networks will improve the interface quality in vertically stacked devices (53). To highlight the importance of network morphology in vertically stacked devices, LPE graphene networks of two different nanosheet lengths ($l_{NS}$ = 630 nm and 215 nm) were coated with silver nanoparticles (AgNP, diameter ~50 nm), mimicking top electrode deposition. 3D-imaging (Fig. 5B) could resolve individual nanoparticles, which penetrated ~1.3 μm into the network of larger nanosheets, but not into the network of smaller nanosheets. This aligns with Fig. 3B-F, which shows networks of smaller nanosheets to be more densely packed with more tortuous pore volumes. To interrogate the AgNP/graphene interface, we measured their respective volume fractions as a function of depth in the out-of-plane (y) direction from the top surface of each heterostack (Fig. 5B). By using smaller nanosheets in printed LPE networks, the degree of interlayer penetration can be dramatically reduced, avoiding shorting and leading to improved performance in printed TFTs and capacitors (54, 55).

Moreover, this technique can be used to 3D-image complex devices, differentiating various components. Shown in Fig. 5C is a 3D-image of an ITO/Glass/$WS_2$/evaporated-Au vertical heterostack. Four-way segmentation allowed this device to be separated into its discrete layers to enable analysis of the internal nanostructure. By removing the $WS_2$ and pores, we could visualise the relationship between substrate and gold, identifying electrical shorts between top and bottom electrodes (56). Also, discrete device layers like the Au top electrode can be isolated and analysed individually (Fig. 5C). Here, the roughness of the underlying $WS_2$ network has caused holes to form, leading to a poor-quality gold film, as reflected by $\kappa_{IP}$ = 1.8, implying electrode resistances roughly double what might be expected.

In this work, experimental constraints limit us to non-cubic (5 nm × 5 nm × 15 nm) voxels, which limits resolution and hinders analysis. Although one can produce cubic voxels by linear interpolation between adjacent frames (*57*), this yields image elements with blurred edges. Faced with similar problems, the computer-vision community utilise neural networks, such as DAIN (*58*). These algorithms, usually trained on the Vimeo90K dataset (*59*), generate additional frames between consecutive images, increasing the resolution along the time direction. Here, this strategy can improve the resolution along the cutting direction of FIB-SEM-NT data. To test this approach, frames were removed from one image stack and replaced by images generated by DAIN. These can then be compared to the removed ground-truth frames, as displayed in Fig. 6A-C. We find extremely good agreement showing that neural-network based approaches can be used to enhance resolution in FIB-SEM-NT generated 3D images.

## **Conclusion**

3D-imaging using FIB-SEM-NT allows us to quantify the morphology of nanostructured networks, yielding a number of important parameters. This approach is versatile and can be applied to networks of 1D and 2D nanomaterials which are both conducting and semiconducting. Multi-phase segmentation allows FIB-SEM-NT to be extended and applied to heterostacks and devices. We believe this technique will be an important tool to investigate a range of nano-enabled devices.

## **Acknowledgements**

We acknowledge the European Research Council Advanced Grant (FUTURE-PRINT) and the European Union under Graphene Flagship core 3 (grant agreement 881603). We greatly appreciate generous support from the Science Foundation Ireland (SFI) funded centre AMBER (SFI/12/RC/2278) and have availed of the facilities of the SFI-funded AML and ARL labs. We thank Dr. Megan Canavan and Mr. Clive Dowling for valuable help and advice with the FIB-SEM-NT. LD thanks the SFI-funded Centre for Doctoral Training in Advanced Characterisation of Materials (award number 18/EPSRC-CDT/3581) for support. LJ is supported by SFI award URF/RI/191637.

# Figures

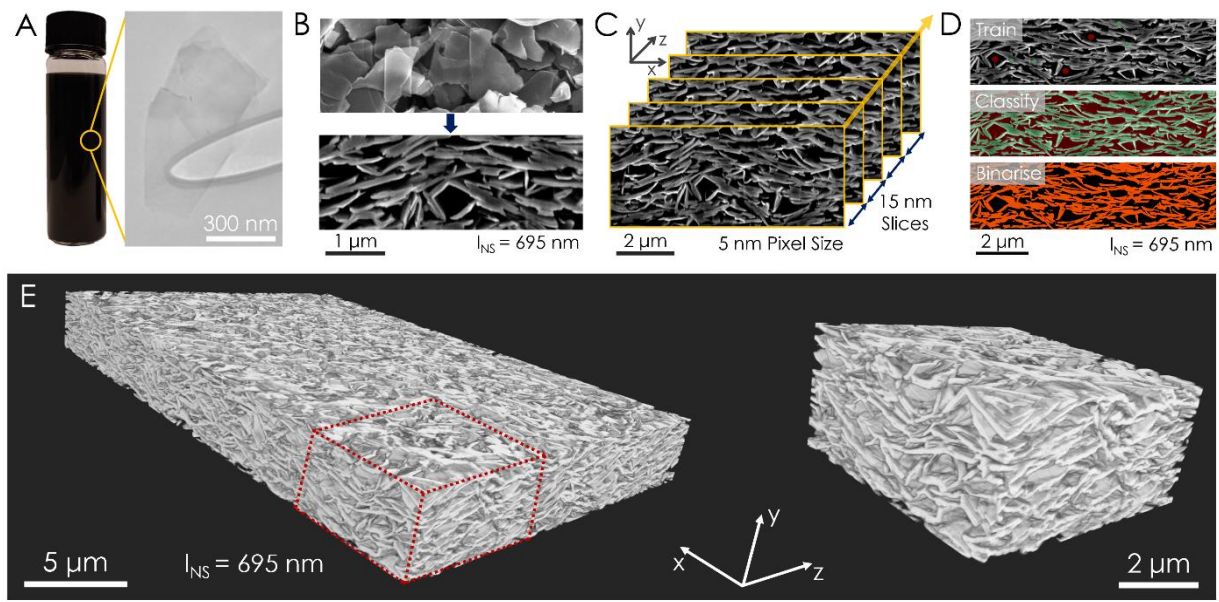

**Fig. 1 FIB-SEM Nanotomography of a Printed LPE Graphene Network**

**(A)** Photo of a graphene dispersion and typical TEM image of a liquid-exfoliated nanosheet. **(B)** Representative surface and cross-sectional SEM images of a printed multilayer graphene nanosheet network (nanosheet length, $l_{NS}$ = 695 nm). **(C)** Schematic of the slice-and-scan process. Network cross-sections are sequentially milled and imaged to produce a stack of 15 nm thick slices. **(D)** Image segmentation pipeline to classify a greyscale network cross-section into its nanosheet and pore components. **(E)** 3D reconstruction of a printed LPE graphene network generated using FIB-SEM-NT. Inset: Magnified region showing nanosheets and pores.

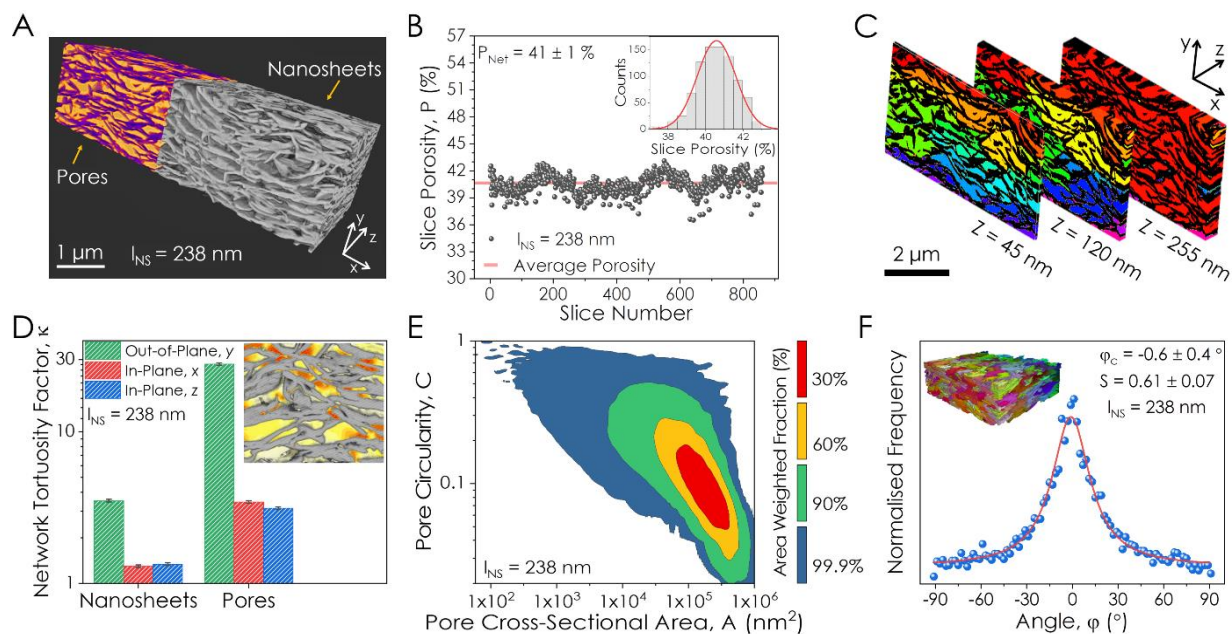

**Fig. 2 Quantitative Analysis of a Reconstructed LPE Graphene Network**

**(A)** 3D reconstruction of a printed graphene network ($l_{NS}$ = 238 nm) separated into its pore and nanosheet contributions. **(B)** Porosity measurements for each slice in the network. The red line is the average porosity across the 3D volume. Inset: Histogram of slice porosity values. **(C)** Pore connectivity analysis for reduced network volumes of thickness z = 45, 120 and 255 nm. As the thickness of the interrogated volume increases, the colour-coded discrete pores coalesce into a connected (red) macropore. Nanosheets are coloured black. **(D)** Tortuosity factor measurements in the out-of-plane (*y*) and in-plane (*x,z*) directions for the nanosheet and pore volumes. Inset: Heat map of diffusive flux in the z-direction through the network. **(E)** Pore circularity plotted as a function of cross-sectional area, A, for each pore chamber in the network. The contours denote the percentage of total area contained within each band. **(F)** Distribution of angles (φ) between the nanosheet normal vectors and the out-of-plane (*y*) direction in the network. The solid line is a fit to a Cauchy-Lorentz distribution centred on φ$_C$ = -0.6°. Inset: Discrete 2D objects in the network isolated using a 3D distance transform watershed.

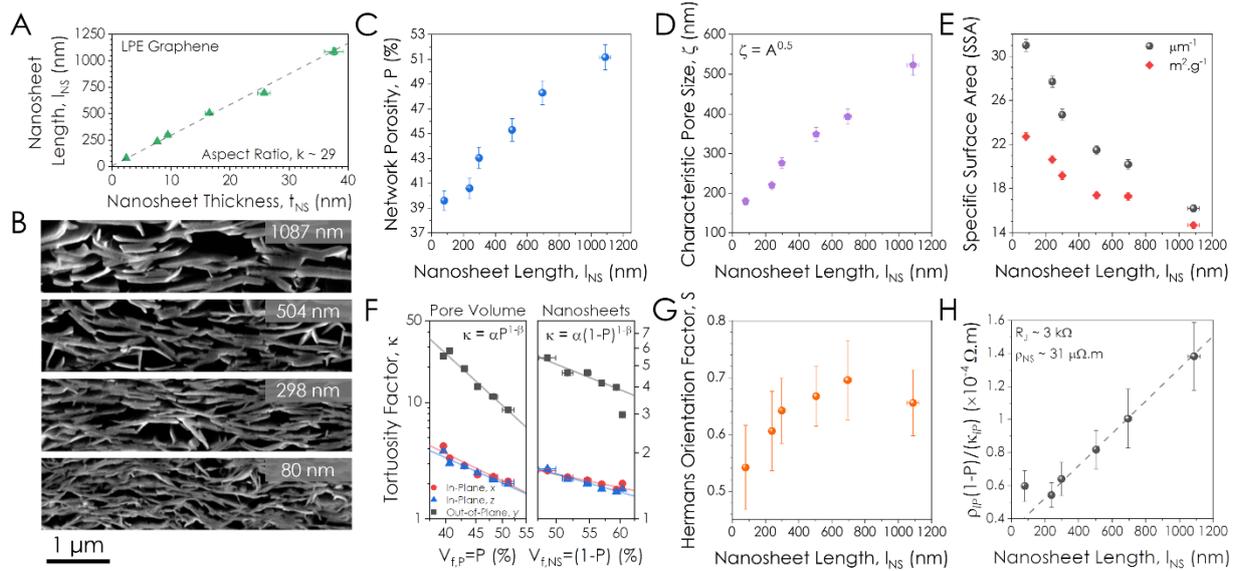

**Fig. 3 Morphology of Printed Graphene Networks as a Function of Nanosheet Size**

**(A)** Relationship between the mean nanosheet length, $l_{NS}$, and thickness, $t_{NS}$, for each 2D ink. **(B)** Representative FIB-SEM cross-sections for different nanosheet lengths. **(C)** Plot of network porosity, P, as a function of $l_{NS}$. **(D)** Scaling of the mean pore size, $\zeta = \sqrt{\langle A \rangle}$, in each network with $l_{NS}$. **(E)** Network specific surface area plotted against $l_{NS}$ in (µm$^2$·µm$^{-3}$) and (m$^2$·g$^{-1}$) units. **(F)** Plot of the pore and nanosheet tortuosity factors in the out-of-plane (y) and in-plane (x,z) directions as a function of volume fraction of pores (P) and nanosheets (1-P) respectively. The solid lines are fits to an adjusted Bruggeman relation described by $\kappa = \alpha P^{1-\beta}$ for the pore data and $\kappa = \alpha(1-P)^{1-\beta}$ for the nanosheets, where $\alpha$ is a prefactor and $\beta$ is the fitted Bruggeman exponent. **(H)** Hermans orientation factor, S, plotted as a function of nanosheet length. **(I)** Plot of the morphologically scaled network resistivity $(\rho_{IP}(1-P))/\kappa_{IP}$ against $l_{NS}$. The in-plane electrical resistivity is described by $\rho_{IP}$ and $\kappa_{IP}$ is the in-plane tortuosity factor of the nanosheets. The straight line is a fit to Eqn. 1.

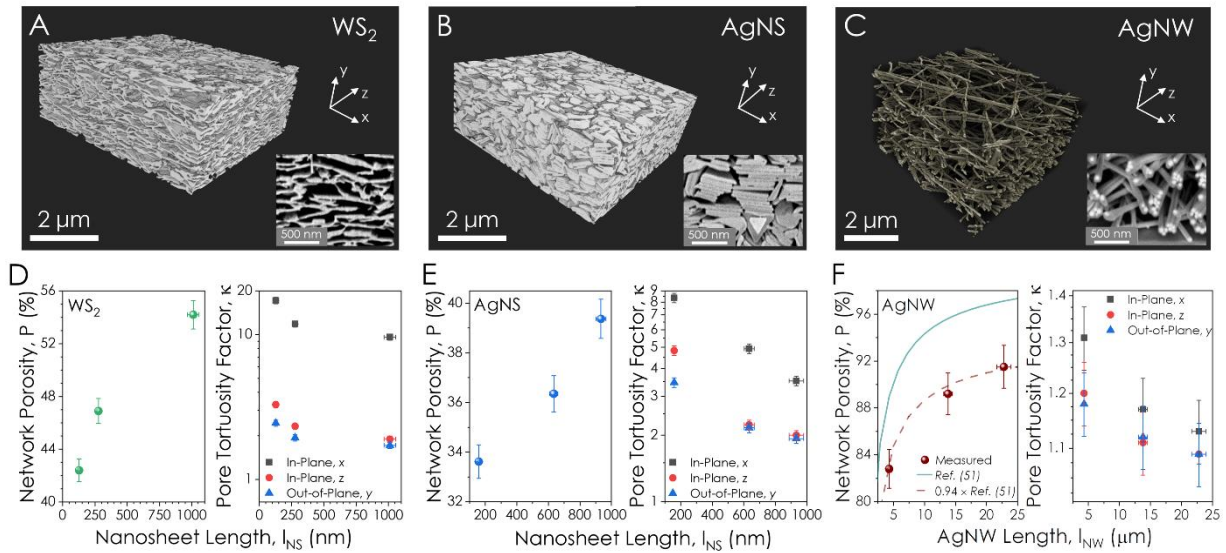

**Fig. 4 Morphological Characterisation of Materials Beyond Graphene**

Reconstructed network volumes of printed **(A)** WS$_2$ nanosheets, **(B)** Silver nanosheets (AgNS) and **(C)** Silver nanowires (AgNWs). Inset: Representative FIB-SEM cross-sections for each material. **(D-F)** Plots of the network porosity and pore tortuosity factor as a function of nanosheet/nanowire length for the **(D)** WS$_2$, **(E)** AgNS and **(F)** AgNW networks. The solid line in **(F)** is the predicted porosity scaling with AgNW length (*51*), which we find to describe our data well when scaled by a prefactor of 0.94 (dashed line).

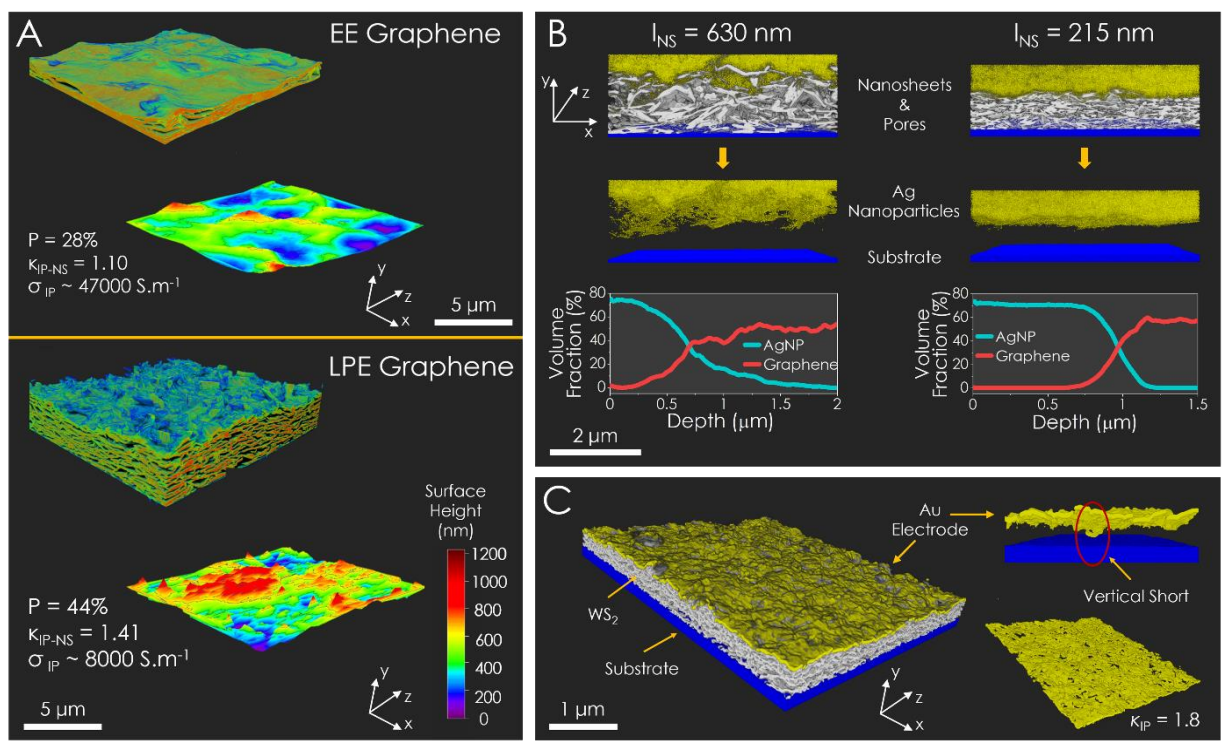

**Fig. 5 Characterisation of Electrochemically Exfoliated (EE) Graphene Networks, Vertical Heterostacks and Nanostructured Devices**

**(A)** 3D reconstructions and surface topography of printed LPE and EE graphene networks to compare the network morphologies. **(B)** Printed graphene/silver nanoparticle stacks for two different nanosheet sizes ($l_{NS}$ = 215 and 630 nm) to characterise the degree of interlayer penetration with changing $l_{NS}$. The volume fraction of both phases is plotted as a function of depth in the out-of-plane (*y*) direction from the top surface of each heterostack. **(C)** A reconstructed Glass/ITO – $WS_2$ – Evaporated Au device segmented into its discrete layers. Removal of the $WS_2$ mid-layer allows vertical shorts through the $WS_2$ network to be identified. The Au electrode has been isolated to show the presence of holes in the layer.

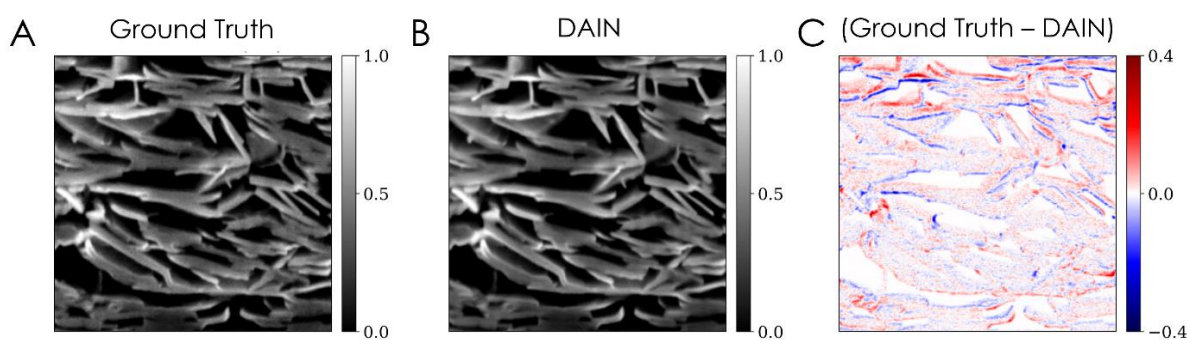

**Fig. 6 Computer Generated Intermediate Images**

**(A)** One of the ground-truth images removed from the original stack of FIB-SEM data. **(B)** The correspondent image generated by the video frame interpolation algorithm DAIN (*58*). **(C)** The difference between these two images.